\documentclass[aps,prx,twocolumn,amsmath,amssymb]{revtex4-1}
\usepackage{graphicx}
\usepackage{dcolumn}
\usepackage{bm}
\usepackage{amsmath}
\usepackage{amsfonts}
\usepackage{amssymb}
\usepackage{multirow}

\usepackage{amsthm}
\usepackage{pinlabel}
\usepackage{natbib}

\usepackage{epstopdf}
\usepackage[abs]{overpic}

\usepackage[dvipsnames]{xcolor}

\newcommand{\no}{\nonumber}

\usepackage{braket}

\definecolor{specialgray}{HTML}{505050}
\definecolor{col10K}{HTML}{FFA000}
\definecolor{col300K}{HTML}{924FA4}
\definecolor{colMu}{HTML}{5278BD}
\definecolor{colMuI}{HTML}{924FA4}

\begin{document}

\title{Self-consistent temperature dependence of quasiparticle bands in monolayer FeSe on SrTiO$_3$}
\author{Fabian Schrodi}\email{fabian.schrodi@physics.uu.se}
\author{Alex Aperis}\email{alex.aperis@physics.uu.se}
\author{Peter M. Oppeneer}\email{peter.oppeneer@physics.uu.se}
\affiliation{Department of Physics and Astronomy, Uppsala University, P.\ O.\ Box 516, SE-75120 Uppsala, Sweden}

\vskip 0.4cm
\date{\today}

\begin{abstract}
\noindent 
We study the temperature evolution of the quasiparticle bands of the FeSe monolayer on the SrTiO$_3$ (STO) substrate from 10 to 300\,K by applying the anisotropic, multiband and full-bandwidth Eliashberg theory. To achieve this, we extend this theory by self-consistently coupling the chemical potential to the full set of Eliashberg equations. In this way, the electron filling can accurately be kept at a constant level at any temperature. Solving the coupled equations self-consistently, and with focus on the interfacial electron-phonon coupling, we compute a nearly constant Fermi surface with respect to temperature and predict a non-trivial temperature evolution of the global chemical potential. This evolution includes a total shift of 5\,meV when increasing temperature from 10 to 300\,K and a hump-like dependence followed by a kink at the critical temperature $T_c$. We argue that the latter behavior indicates that superconductivity in FeSe/SrTiO$_3$ is near to the BCS-BEC crossover regime. Calculating  the temperature dependent Angle Resolved Photoemission Spectroscopy (ARPES) spectra, we suggest a new route to determine the energy scale of the interfacial phonon mode by measuring the energy position of second-order replica bands. Further, we re-examine the often used symmetrization procedure applied to such ARPES curves and demonstrate substantial asymmetric deviations. Lastly, our results reveal important aspects for the experimental determination of the momentum anisotropy of the superconducting gap.
\end{abstract}

\maketitle

\section{Introduction}

The iron selenide (FeSe) monolayer grown on strontium titanate (STO) shows superconductivity at an extremely high critical temperature of $T_c\sim50-70$\,K \cite{Lee2014, Zhang2017, Liu2012, Qing-Yan2012, He2013, Tan2013}, in stark contrast to the bulk FeSe value of around $8$\,K \cite{Hsu2008}. This observation has led to a huge interest in probing the increase in $T_c$ caused by few-layer materials grown on a substrate \cite{Logvenov2009,Miyata2015,Wang2016d}. One of the important observations of various ARPES experiments is the appearance of replica bands, and the rather strong dependence of the critical temperature and other characteristic experimental results on the electron doping of the system \cite{Wen2016, Lei2016}. 
On the theory side, it was suggested \cite{Lee2014,Rademaker2016} and has been shown recently via calculations specific to FeSe/STO \cite{Aperis2018}, that the high transition temperature in the single-layer can be explained by a small-${\bf q}$ electron-phonon interaction (EPI) that arises at the interface. Antiferromagnetic spin fluctuations have been predicted for bulk FeSe \cite{Scherer2017} and very recently, for monolayer FeSe on STO \cite{Shishidou2018}, yet their role for the superconductivity still needs to be clarified. There is, moreover, a lack of a completely self-consistent, temperature-dependent theory, that not only explains the change of results in ARPES experiments with $T$ \cite{Qing-Yan2012, Tan2013}, but is in addition capable of making predictions for the temperature evolution of not yet measured, though resolvable, quantities like the global chemical potential. This  quantity was measured e.g.\ for bulk FeSe and found to have a totally non-trivial behavior \cite{Rhodes2017}.

Here we present the first temperature dependent, full bandwidth, multiband and anisotropic Eliashberg theory extended with an additional equation, that self-consistently keeps the electron filling constant. This, in turn, ensures not to change the properties of the system due to electron doping, as we raise the temperature. Starting from the observations of Ref.\,\cite{Aperis2018} we introduce a small-momentum electron-phonon coupling as the superconductivity mediating mechanism. After self-consistently solving the extended set of Eliashberg equations we calculate a shift of the global chemical potential of $\sim5$\,meV when going from $10$\,K to $300$\,K, as well as a hump-like behavior below $T_c$. Such behavior is characteristic of systems with large gaps and shallow bands \cite{Marel1990,Rietveld1990} and indicates that superconductivity in FeSe/STO is near to a BCS-BEC crossover, similar to bulk FeSe \cite{Lubashevsky2012}. We observe no significant temperature-induced changes in the positions of either the main band at $\sim-50$\,meV or the replica band at $\sim-160$\,meV. We also find weak second-order replica bands whose peak position lies below the main replica bands at an energy that equals exactly the characteristic frequency of the interfacial phonon. This energy difference does not depend on the electron-phonon coupling strength, in contrast to the location of the main replica band \cite{Rademaker2016,Aperis2018}. Thus, we suggest that the detection of the weaker replica bands can provide a definite measure of the energy scale of the involved interfacial phonon mode. Our results show that an additional feature appears with increasing $T$ in the ARPES spectrum at zero energy. This peak originates from thermal broadening effects of the electron-boson interaction at the Fermi level, spreading out and transferring spectral weight to the $M$ point. Further we test explicitly the symmetrization procedure, which is a commonly used procedure in the evaluation of ARPES data \cite{Lee2014, He2013}, and find a non-negligible deviation in the spectral function, compared to the non-symmetrized results. By mimicking the superconducting gap measurement procedure usually applied in experiment, we report a significant sensitivity of the momentum dependence with respect to the measurement angle and the Fermi surface sampling. As a consequence, we find that the location of the observed gap maxima can change from being at the intersection of the two elliptical electron Fermi sheets to being along the major axis of the ellipsis. The latter anisotropy agrees with recent ARPES observations \cite{Zhang2016}. In addition, for large temperature changes we observe slightly varying results for the Fermi surface. The chemical potential renormalization average over momenta is found to be nearly constant with respect to temperature, though still a function of energies, while the exact reverse is true for the global chemical potential $\mu$. Regarding momenta on the Fermi surface, there are clear signatures of this renormalization to become more isotropic with raising $T$, developing the tendency for an increasingly global competition with $\mu$ at the Fermi level.

\section{Methodology}

We build upon the theory of Ref.\,\cite{Aperis2018}, which has shown the crucial importance of the electron-phonon interaction to account for the high $T_c$ observed in experiments. It was revealed that the influence of so-called deep Fermi sea Cooper pairing is non-negligible, making a multi-band treatment, that includes also the bands not crossing the Fermi level, a necessity \cite{Aperis2018}. 
Within this treatment the electron density was kept fixed by adjusting the chemical potential so as to satisfy the respective equation for the electron filling \cite{Aperis2018}. However, this procedure makes it difficult to efficiently account, with the needed precision, for changes in the system's chemical potential, e.g. when the temperature is varied, so as to accurately predict the concomitant temperature evolution of the quasiparticle spectra.
This is why we increase here the number of coupled equations within Eliashberg theory by one, explicitly including the calculation of the chemical potential in a self-consistent manner. By doing so, we not only prevent the electron density from changing, but we are also able to determine variations in the global chemical potential up to numerical accuracy.

The electron-phonon interaction is modeled by small-${\bf q}$ phonons derived from the isotropic mode $\hbar\Omega=81$\,meV of the interface \cite{Lee2014,Xie2015}. The FeSe electrons are coupled to these phonons by $g({\bf q})=g_0\exp(-|{\bf q}|/q_c)$, where $g_0$ is the global effective electron-phonon scattering strength, $q_c=0.3a^{-1}$, and $a$ is the FeSe lattice constant. The coupling constant $g_0=728$\,meV is found by imposing the position of the main energy band at the $M$ point of the folded Brillouin zone (BZ) to be at around $-50$\,meV and the replica band to appear at $-160$\,meV \cite{Aperis2018}, which are the values observed by ARPES measurements \cite{Lee2014, Zhang2017}. To match the experiment as reliable as possible, we take the experimental temperature $T=10$\,K for calculating $g_0$. In this way we take electron screening effects implicitly into account and do not have to include them in our Hamiltonian, which is shown in Eq.\,(\ref{hamiltonian}) in Appendix A. 

We use a ten-band tight-binding energy dispersion of bulk FeSe, as developed in Ref.\,\cite{Eschrig2009} by a fit to Density Functional Theory (DFT) calculations, and modified with respect to the relevant hopping parameters, to account for the monolayer situation, in Ref.\,\cite{Hao2014}. For a given temperature $T$ and initial choice of the chemical potential $\mu^{(I)}$, which rigidly shifts the momentum 
(${\bf k}$) and band ($n$) dependent bare energy dispersion $\xi_n^b({\bf k})$, the electron filling of a system with $L$ bands in the normal state is given by
\begin{eqnarray}
n_0 &=& 1 + \frac{2T}{L}\sum_{{\bf k}^{\prime},m^{\prime}}\sum_n \frac{\xi_n^b({\bf k}^{\prime})-\mu^{(I)}}{\omega_{m^{\prime}}^2+\left[\xi_n^b({\bf k}^{\prime})-\mu^{(I)}\right]^2} .
\label{n0}
\end{eqnarray}
This expression is derived from a non-interacting theory in Matsubara space, where $\omega_m=\pi T(2m+1)$ are the fermionic frequencies. 
Since the electron filling is to be kept constant at a particular value $n_0$, to model an experimental situation where the temperature is varied but without moving charges, we can invert Eq.\,(\ref{n0}) to find self-consistently the associated chemical potential. In this simple case of the normal state, the infinite Matsubara summation can be taken care of analytically, yielding the following expression,
\begin{widetext}
	\begin{eqnarray}
	\mu^{(I)} &=& \left[\sum_{{\bf k}^{\prime},n}\frac{\xi_n^b({\bf k}^{\prime})}{\xi_n^b({\bf k}^{\prime})-\mu^{(I)}}\tanh\left(\frac{ \xi_n^b({\bf k}^{\prime})-\mu^{(I)}}{2T}\right) + (1-n_0)L   \right]  \cdot \left[ \sum_{{\bf k}^{\prime},n}\frac{1}{\xi_n^b({\bf k}^{\prime})-\mu^{(I)}}\tanh\left(\frac{ \xi_n^b({\bf k}^{\prime})-\mu^{(I)}}{2T}\right) \right]^{-1} ~~
	\label{muInit}
	\end{eqnarray}
\end{widetext}
which can be calculated straight-forwardly. 

Our numerical tests reveal that this equation can be implemented in a robust way only by introducing the hyperbolic tangents, i.e., by making use of the infinite summation instead of a finite interval. Armed with Eq.\,(\ref{muInit}), giving a chemical potential that corresponds to the desired electron filling, we follow the Eliashberg treatment to find three coupled equations for the mass renormalization function $Z$, the chemical potential renormalization $\chi$ and the superconducting gap function $\phi$ when the electron-boson interaction is turned on. The expressions for these quantities are given in Eqs.\,(\ref{coupledfun1})-(\ref{coupledfun3}). The electron filling within this formalism changes to
\begin{eqnarray}
n_1 &=& 1 - \frac{2T}{L}\sum_{{\bf k}^{\prime},m^{\prime}}\sum_n \frac{\xi_n^b({\bf k}^{\prime})-\mu+\chi({\bf k}^{\prime},i\omega_{m^{\prime}})}{\Theta_n({\bf k}^{\prime},i\omega_{m^{\prime}})} ~~,
\label{n1}
\end{eqnarray}
where $\Theta_n$ is given by Eq.\,(\ref{couplednorm}), and we impose that $n_1=n_0$, while in general $\mu\neq\mu^{(I)}$. Inverting Eq.\,(\ref{n1}) is less straight forward than treating the bare case, since the Matsubara summation cannot be directly evaluated. We therefore make the following assumption: The summand in Eq.\,(\ref{n1}) can accurately be approximated by the normal-state expression for any $\omega_m$ above a threshold $\mathcal{M}$, i.e., for $|m|>\mathcal{M}$.
Using this assumption we find the chemical potential $\mu$ which ensures a constant electron density, and is strongly coupled to the functions $Z$, $\chi$ and $\phi$, which again are functions of $\mu$, as follows 
\begin{widetext}
	\begin{eqnarray}
	\mu &=&\left[\frac{1}{2T}\sum_{{\bf k}^{\prime},n}\left(\tanh\left(\frac{ \xi_n^b({\bf k}^{\prime})-\mu^{(I)}}{2T}\right) - \tanh\left(\frac{ \xi_n^b({\bf k}^{\prime})-\mu}{2T}\right)\right) + \sum_{{\bf k}^{\prime},n}\sum_{|m^{\prime}|\leq \mathcal{M}} \left( \frac{\xi^b_n({\bf k}^{\prime}) + \chi({\bf k}^{\prime}, i\omega_{m^{\prime}})}{\Theta_n({\bf k}^{\prime}, i\omega_{m^{\prime}})}  \right. \right.  \nonumber \\
	&& ~~~~~~~~~~~~ ~~~~~\left. \left. + \frac{\xi^b_n({\bf k}^{\prime})}{\omega_{m^{\prime}}^2+\left[\xi_n^b({\bf k}^{\prime})-\mu\right]^2}  \right) \right] \cdot \left[ \sum_{{\bf k}^{\prime},n}\sum_{|m^{\prime}|\leq \mathcal{M}} \left( \frac{1}{\Theta_n({\bf k}^{\prime}, i\omega_{m^{\prime}})} + \frac{1}{\omega_{m^{\prime}}^2+\left[\xi_n^b({\bf k}^{\prime})-\mu\right]^2}  \right) \right]^{-1}  ~~
	\label{mu}
	\end{eqnarray}
\end{widetext}
We note that this result is invariant under any shift of $\mathcal{M}$, as long as the corresponding assumption is fulfilled. 
Again, the inclusion of the infinite Matsubara frequency terms is necessary to ensure the stability and the reliable convergence of our algorithm. 
The expression for the chemical potential has been implemented in the Uppsala Superconductivity (UppSC) code \cite{Aperis2015,Bekaert2018,Aperis2018,UppSC}, which solves self-consistently the Eliashberg equations on the basis of \textit{ab initio} calculated input.
To our knowledge, we are the first to solve this set of four highly coupled equations iteratively and obtain full bandwidth, momentum, temperature, and energy dependent results after analytic continuation from the imaginary to the real axis (see Appendices \ref{app1} and \ref{app2} for details).

\section{Results}

\begin{figure}[h!]
\includegraphics{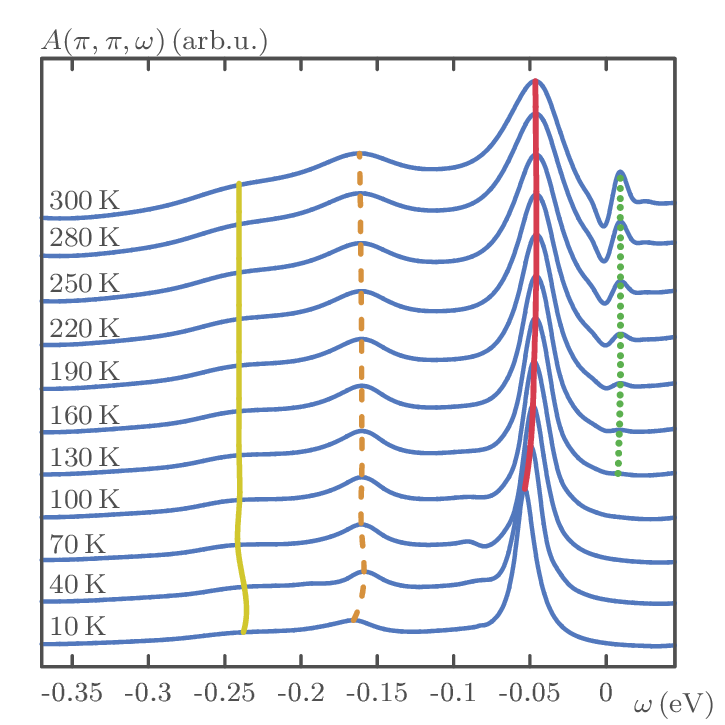}
\caption{Self-consistently calculated spectral function at the high-symmetry $M$-point of the folded Brillouin zone for different temperatures. The main peak lies at approximately $-50$\,meV, its position changes slightly as temperature increases (solid red line). The dashed orange line shows the maximum of the replica band which is peaked near $-160$\,meV, well in agreement with experiment \cite{Lee2014,Zhang2017}. We observe also a second-order replica band, depicted by the yellow line, at around $-240$\,meV. At temperatures above $100$\,K a feature at slightly positive energies emerges due to thermal broadening effects (dotted green line). Just as we find for the position of the main band, the energy positions of all replica bands barely move as the temperature increases.} 
\label{arpes}
\end{figure}

Within the theory that we present here, the momentum, temperature and energy dependent spectral function can easily be obtained from the Green's function. In Fig.\,\ref{arpes} we show the temperature evolution of the spectral function $A({\bf k},\omega)$ evaluated at the $M$ [=($\pi$,$\pi$)] point of the folded Brillouin zone, corresponding to the experimentally relevant situation. Such spectra correspond to the so-called energy distribution curves (EDCs). The maximum value of the main band that changes slightly with increasing temperature, is drawn in red as guide for the eyes. The replica band (denoted in orange, dashed) is located at energies around $-160$\,meV . Its energy position does not show any significant $T$ dependence. The same observation holds true for the higher order replica band, shown in yellow in Fig.\,\ref{arpes}, that we are able to resolve within our calculations. Such a second order replica peak was previously reported within one-band model calculations \cite{Rademaker2016}. Its appearance is therefore a generic feature of the interfacial small-$\bf q$ EPI, independent of details of the replicated electron bands around the $M$ point. Above $T=100$\,K still another, and yet unexpected feature, highlighted by the dotted green line on the right-hand side, forms at energies very slightly above zero. Further, we find a non-negligible broadening, due to thermal effects, of all peaks below $\omega=0$\,eV for increasing temperatures.

At this point it deserves to be mentioned that ARPES experiments \cite{Lee2014,Rebec2017,Zhang2017} have detected the electron band at the $M$ point and also a deeper lying hole band at the same position in momentum space. The latter hole band is not present in the here-used tight-binding energy dispersions for monolayer FeSe (see Refs.\ \cite{Hao2014,Aperis2018}). The measured band dispersions of monolayer FeSe could thus far not be reproduced by correlated band-theory calculations as dynamical mean field theory (DMFT) \cite{Mandal2017,Nekrasov2018}. As the here-used tight-binding bands provide a very good description of the bands in the near-Fermi energy region we therefore preferred to use these.

Despite the fact that we can reproduce the observed striking features of the electron band detected in ARPES experiments very well, there are as yet no direct measurements of the second-order replica band, nor of the temperature dependent feature evolving at positive energies. 
Concerning the main replica band near $-160$ meV, it deserves to be mentioned that a recent work  attributes the appearance of this band to the interaction of the outgoing photoelectron with a surface phonon \cite{Li2018}. Here, however, we compute it from the interfacial electron-phonon interaction of FeSe electrons with the substrate phonon. A proof of this mechanism would be the observation of the here-predicted second-order replica band.
Its  thus far lacking detection can be explained by limited ARPES resolution, since the signal is expected to be very weak.

The binding energy difference between the location of the main and our second-order replica band is equal to the characteristic energy of the interfacial phonon used in our theory, $\Omega=81$\,meV. However, the energy distance between our obtained main replica bands and the electron bands that form the Fermi surface around $M$ is significantly larger than $\Omega$ \cite{Aperis2018}. This is in contrast to previous findings where both first and second order  replica bands were found to appear always in multiples of $\Omega$ below the main electron bands \cite{Rademaker2016}. The reason for this difference may be the significantly lower values of the coupling strength needed to explain the $T_c$ in Ref.\,\cite{Rademaker2016} and/or the simplicity of the effective model used.
In fact, one can qualitatively show that, while the distance between the main bands and the first-order replicas depends on both $\Omega$ and the coupling strength \cite{Rademaker2016,Aperis2018}, the relative difference in energy between one replica band and its next order counterpart is approximately equal to the frequency of the involved phonon mode regardless of the coupling strength.
Therefore, measuring the energy location of the second-order replica band with ARPES in FeSe/STO can provide valuable insight not only on the characteristic energy scale of the interfacial phonon mode but also on the overall coupling strength when combined with a measurement of the main replica bands.

For the detection of the positive-energy peak which we predict in Fig.\,\ref{arpes}, there are two experimental difficulties to be overcome: obtaining ARPES data at positive energies (although only very slightly above zero) and ensuring not to damage the sample at higher temperatures. The origin of this feature is the thermal broadening of quasiparticle occupancies at the Fermi level in combination with the electron-phonon interaction that spreads the spectral weight across the Brillouin zone. In other words, the dip at exactly zero frequency resembles the properties of a Fermi surface point nearby. Regarding the position of the main bands, measurements in bulk FeSe reveal a shift toward higher binding energies with increasing temperature \cite{Rhodes2017}, a trend we do not find for the single layer case. 

\begin{figure}[ht!]
\includegraphics[width = 1.0\columnwidth, clip, unit=1pt]{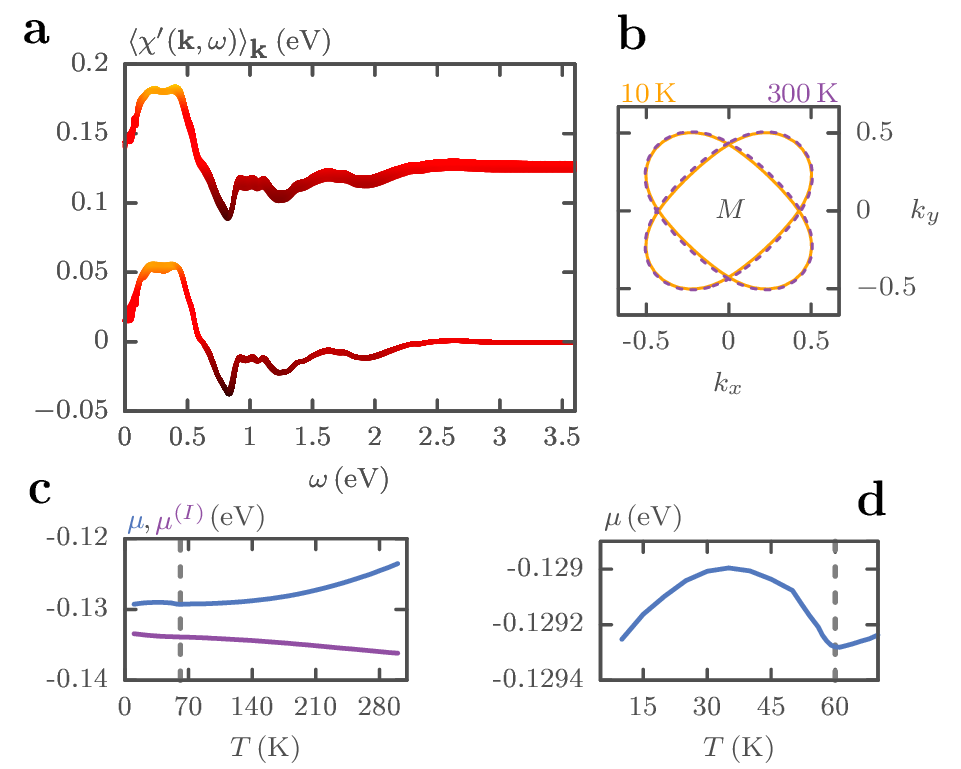}
\caption{{(a)} Calculated energy-dependent real part of the chemical potential renormalization, plotted for several temperatures in the range $10\,\text{K}\leq T\leq300\,\text{K}$; the lower curve represents the self-consistent result for $\langle \chi^{\prime}({\bf k},\omega)\rangle_{\bf k}$, the upper curve includes the global shift due to $\mu$, i.e., $\langle\chi'({\bf k},\omega)\rangle_{\bf k}-\mu$. {(b)} Fermi surface for $T=10$\,K (orange) and $T=300$\,K (purple). {(c)} Comparison between the normal, non-interacting state chemical potential $\mu^{(I)}$, shown in purple, and the self-consistent result $\mu$, depicted in blue, as a function of temperature. {(d)} Computed results for $\mu$ within a temperature interval with upper bound slightly above $T_c$ (indicated by the gray dashed line), which shows a well-pronounced hump and a kink at the transition temperature.}
\vspace*{-0.3cm}
\label{chemPotTDep}
\end{figure}

Turning to the temperature dependent evolution of the Fermi surface, it has been shown that the iron-based superconductors show a quite large tendency for rigid band shifts due to the global chemical potential \cite{Brouet2013, Dhaka2013}. A possible explanation for this is the shallowness of the electron and hole pockets \cite{Brouet2013}. Since recent experiments on bulk FeSe have shown a $10$\,meV change in the chemical potential \cite{Rhodes2017}, it is worth examining whether this trend applies for the monolayer as well. As is evident from Eq.\,(\ref{mu}) our theory \textit{self-consistently} allows for such rigid shifts with temperature, which are supplementary to the chemical potential renormalization function $\chi$. In Fig.\,\ref{chemPotTDep}{(c)} and {(d)} the evolution of $\mu$ can be seen to be non-trivial in the range of $10\,\text{K}\leq T\leq300\,\text{K}$. For sake of comparison we also plot the normal-state behavior calculated from Eq.\,(\ref{muInit}) (purple line in Fig.\,\ref{chemPotTDep}{(c)}) that shows an opposite trend. Since up to this date there are no corresponding measurements of the chemical potential, we predict not only a change of $\sim5$\,meV in this experimentally resolvable temperature range, but also a hump-like shape for $T<T_c$. Our predicted shift is large enough to be non-trivial, but too small to introduce a topological change in the Fermi surface. This is directly revealed in Fig.\,\ref{chemPotTDep}{(b)}, where we show the Fermi surface for $10$\,K and $300$\,K. There are small changes, just large enough to be resolvable, but not of significant size. We note, however, that the deviations are larger for the inner electron pocket, due to the small-${\bf q}$ electron-phonon interaction. 

It would be interesting to separate the effect of thermal broadening from that of the electron-phonon interaction on the calculated $\sim5$\,meV shift in the chemical potential. This may be achieved by comparing the two curves in Fig.\ \ref{chemPotTDep}(c). There, the purple curve is the temperature dependence of the chemical potential ($\mu^{(I)}$) for the non-interacting, non-superconducting system which we can compare with the full interacting result (blue curve) for the temperature range $T_c\leq T\leq300\,\text{K}$. In this temperature interval, the non-interacting $\mu^{(I)}$ varies as $\mu^{(I)}(300\,{\rm K}) - \mu^{(I)}(T_c) = -2.3$ meV, whereas the interacting $\mu$ varies with opposite trend as $\mu(300\,{\rm K}) - \mu(T_c) =+5.8$ meV. The absolute change in the chemical potential due to the electron-phonon interaction is thus 2.5 times larger than what would be expected solely from thermal broadening effects. Yet, if we subtract the purely temperature broadening contribution from our $\mu$, we estimate a chemical potential shift of $\sim8$\,meV when going from $T_c$ to 300\,K caused by interaction effects only.  

\begin{figure}[t!]
\vspace{-0.3cm}
\includegraphics[width = 1.0\columnwidth, clip, unit=1pt]{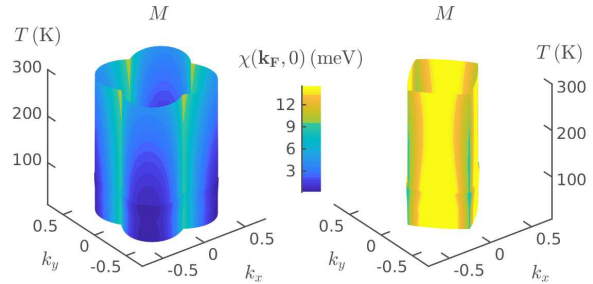}
\vspace{-1cm}
\caption{Calculated temperature evolution for the two Fermi surface sheets, colored with the corresponding zero-frequency value of the chemical potential renormalization function. The phase transition at $T_c$ is reflected in a small kink along the temperature axis.}
\label{3dFSplot}
\end{figure}

Since, in addition to the global $\mu$, there is also the anisotropic renormalization function $\chi$, we show in Fig.\,\ref{chemPotTDep}{(a)} the real part (denoted by a prime) of the ${\bf k}$-averaged value of this quantity. The lower curve corresponds to $\langle\chi^{\prime}({\bf k},\omega)\rangle_{\bf k}$ in the range $10\,\text{K}\leq T\leq300\,\text{K}$  as a function of energy, while the upper curve shows $\langle\chi^{\prime}({\bf k},\omega)\rangle_{\bf k}-\mu$ for the same temperatures. It is nicely seen from Fig.\,\ref{chemPotTDep}{(a)} when comparing the thickness of both curves, that $\mu$ and $\chi$ almost perfectly share the same non-trivial dependencies with respect to energy and temperature. While the global chemical potential is obviously constant with $\omega$, there is almost no change in the Brillouin zone average of $\chi$ when heating the system. On the contrary, $\chi$ exhibits momentum-dependent changes with temperature, as we show in Fig.\,\ref{3dFSplot}. 

In Fig.\ \ref{3dFSplot} we show the temperature evolution of two main energy bands at Fermi surface points ${\bf k}_F$, for temperatures from 10\,K to 300\,\text{K}, colored with the chemical potential renormalization function at frequency $\omega=0$\,eV. From Fig.\,\ref{chemPotTDep}{(b)} we have already seen that there are no large changes in the Fermi surface, neither in the topology, nor in the size. The visualization in Fig.\,\ref{3dFSplot}, however, reveals a small kink at the superconducting transition temperature.
The inner electron band shows lower values of $\chi^{\prime}({\bf k_F},0)$ at the corners and remains essentially unchanged with increasing temperature. More interestingly, the value at the outer energy band not only increases with $T$, but becomes more isotropic due to thermal broadening. This can be understood as $\chi$ developing from a very fine-structured function of momenta and energy, at temperatures not too far above $T_c$, to a more isotropic one for large temperatures that competes more and more globally with the rigid energy shift due to $\mu$.

We now focus on the hump-like shape of our calculated $\mu(T)$ for $T<T_c$. This characteristic behavior is in good agreement with previous BCS mean field calculations where a similar hump-like shape followed by a kink at the transition temperature was found as the ratio $\Delta/\delta\epsilon_F$ approaches one, i.e. as the system approaches the BCS-BEC crossover regime \cite{Marel1990} ($\Delta$ is the BCS superconducting gap and $\delta\epsilon_F$ is the distance of the bottom of the band from the Fermi level). The here-predicted $T$-dependence of $\mu$ could be verified e.g. by work-function measurements \cite{Rietveld1990}. In bulk FeSe, it was demonstrated that this ratio can be as high as 0.5  when the shallow hole bands near the Fermi level are tuned with doping and that this $\Delta/\delta\epsilon_F$ value suffices to drive the system through a BCS-BEC crossover as is evidenced by the non-BCS shape of the Bogoliubov quasiparticle bands seen by ARPES \cite{Rinott2017}.

In FeSe/STO the average value of the superconducting gap near the Fermi level is around 10-15 meV while the bottom of the electron bands at the $M$ point of the BZ lies around 50 meV below the Fermi level, as also found previously \cite{Aperis2018}. This leads to an enhanced $\Delta/\delta\epsilon_F\approx 0.2-0.3$, which places FeSe/STO on the BCS side of the BCS-BEC crossover regime but significantly close to it. Here, despite the fact that we do not observe any deviation from the usual BCS Bogoliubov spectrum as in \cite{Rinott2017},  i.e. our calculated spectral function exhibits the characteristic 'back-bending' near the Fermi level \cite{Aperis2018} as also witnessed experimentally \cite{Lee2014}, our calculated non-trivial dependence of the chemical potential below $T_c$ indicates that FeSe/STO may also be a promising playground to study BCS-BEC crossover phenomena. In contrast to bulk FeSe, in FeSe/STO it is the electron bands that show the BCS-BEC tendency that we find here, simply because the hole bands around $\Gamma$ are far away from the Fermi level as a result of charge transfer at the interface \cite{He2013,Tan2013}. In FeSe/STO, the electron bands are not as shallow as the hole bands of bulk FeSe, however the superconducting gap is much larger than the one observed in the bulk material. It is therefore not peculiar that the monolayer inherits the tendency to BCS-BEC crossover. 

We note that here we focus our discussion on Cooper pair formation at the electron bands. It has recently been shown that due to the large energy scale and the small-$\bf q$ form of the interfacial electron-phonon interaction, pairing at bands away from the Fermi level is also to be expected in FeSe/STO \cite{Aperis2018}. 
In the calculations presented here such deep Fermi sea Cooper pairing is included, however since the binding energy (gap) of this type of pairing is of the order of $\mu$eV, its impact on the spectra that we report should be small. Nevertheless, whether such pairing is of BCS or BEC nature is an interesting open issue.

 It is worth pointing out that, apart from the quantity $\Delta/\delta\epsilon_F$, the evolution from the BCS to the BEC regime is often characterized by the ratio $\xi_0/l$, where $\xi_0$ is the superconducting coherence length and $l$ the interparticle distance (electron mean free path) \cite{Pistolesi1994}. In the extreme BEC limit, $\xi_0\rightarrow 0$ and the electron pairs are tightly bound, thus behaving as bosons. These two ratios are closely related, yet it has been shown that $\Delta/\delta\epsilon_F$ is the best detection parameter of the BCS-BEC crossover regime \cite{Guidini2014}. It is nonetheless customary to provide an estimation of $\xi_0/l$ since its value may provide complementary insights into the BCS-BEC crossover regime, and its calculation is straightforward within the BCS approximation \cite{Pistolesi1994,Guidini2014}. Within Eliashberg theory, the value of this quantity may be accessed by calculating the ratio between the local and London penetration depths \cite{Nam1967}. A discussion on this issue based on single band, isotropic Eliashberg theory is given in Ref.\ \cite{Marsiglio1990} while anisotropic Eliashberg calculations based on \textit{ab initio} input have only just recently become available \cite{Bekaert2016}. In order to provide an estimate of $\xi_0/l$ on the same level of theory as the rest of the calculations presented here, we would need to extend our present Eliashberg theory to include impurity scattering effects which are essential for the proper calculation of the respective penetration depths \cite{Marsiglio1990}. This procedure is out of the scope of the present manuscript, and therefore left for future investigation.

\begin{figure}[ht!]
\includegraphics[width=1.0\columnwidth]{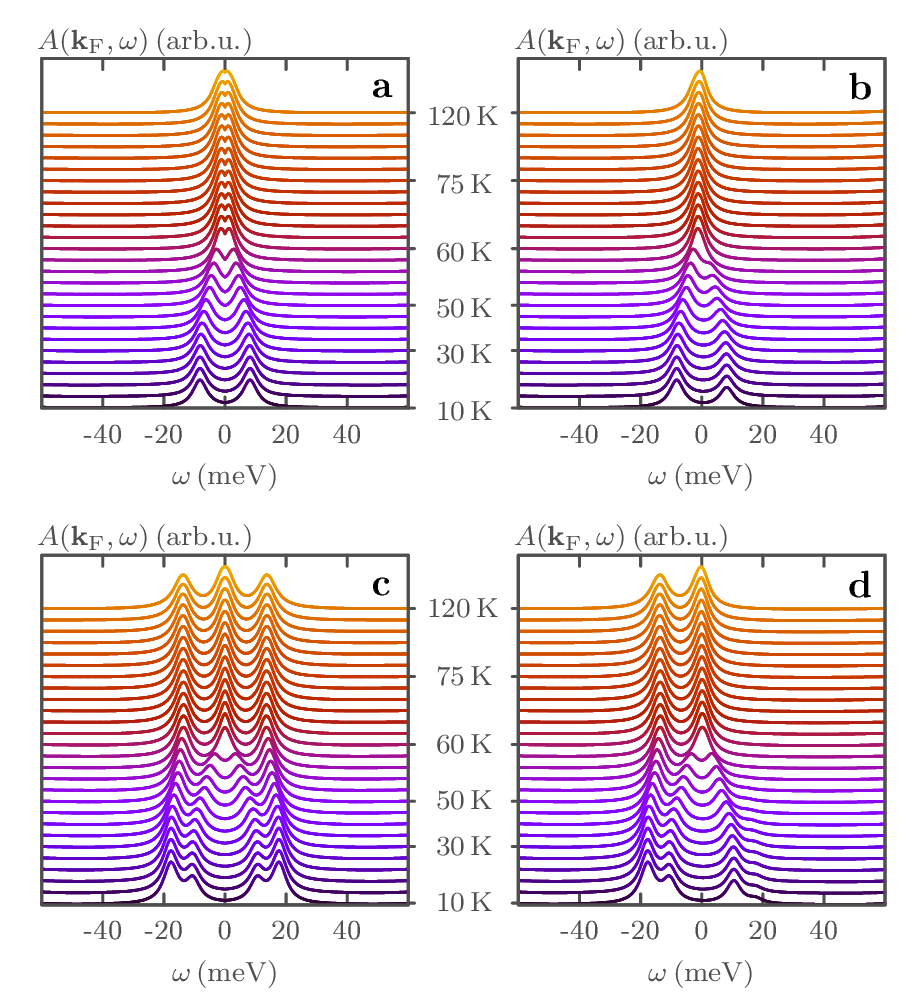}
\caption{{(a)},{(c)} Computed ARPES spectra, symmetrized with respect to zero energy, at two different Fermi surface points; in this way experimental data are usually presented. {(b)},{(d)} Self-consistently computed, non-symmetrized results obtained within our theory. Especially for temperatures above $T_c$ there can be large deviations; note that the symmetrized data in {(c)} yield three intensity maxima at high temperatures, while there are actually only two, as clearly revealed in panel {(d)}.}
\label{symmedc}
\end{figure}

ARPES experiments cannot easily access the $\omega>0$\,eV regime, it is therefore a generally accepted practice to \textit{symmetrize} the measured EDCs with respect to zero energy, and plot the results at a specific Fermi surface point ($k_F$) \cite{Liu2012, Lee2014, Tan2013, He2013}. Since we have access to the full energy range, we plot our results for two different Fermi surface momenta, and compare with the symmetrized versions, in Fig.\,\ref{symmedc}. 
In the upper panels of this plot, we can observe that the gap closing is smoother in the right-hand plot, while in the left (symmetrized) panel there remains a small dip at $\omega=0$\,eV up to high temperatures, which is however completely artificial. The differences are even more drastic in the lower panels of Fig.\,\ref{symmedc}. Starting with the right-hand panel, we see two peaks with energy-symmetric position (but not height) merging into one maximum with increasing temperature and thereby closing the superconducting gap. The leftmost peak changes its position slightly with varying $T$ and remains isolated throughout. This peak is primarily due to crossing of the binding energy of one of the two electron bands that are separated in energy at this specific $k_F$ point. It is therefore less associated to the coherence of Bogoliubov quasiparticles. This explains why the spectral weight of this peak is much weaker at positive energies. Turning to the symmetrized version in Fig.\,\ref{symmedc}{(c)}, not only are the coherence peaks identical in height, but the isolated left maximum is replicated to the right. This results in a physically different situation at high temperatures, namely, that there are three, instead of two maxima. From these observations we learn that the broadly accepted and widely used symmetrization procedure in ARPES experiments should always be handled with caution by physically questioning the genuineness of features appearing in the non-accessible energy range. We report further that the superconducting coherence peaks are always distributed symmetrically around zero energy. By mirroring the data one might in this respect not get the correct peak height at positive energies, but the gap value remains trustworthy. The latter property reflects the particle-hole symmetry of the Bogoliubov spectrum, which of course must be conserved. 
 On the other hand, the peak-height asymmetry around the Fermi level reflects the fact that the underlying normal state system is doped, and therefore, intrinsically particle-hole asymmetric. 
 As a side remark, we note that particle-hole asymmetric ARPES spectra have been discussed before in the context of pseudogap phenomena related to BCS-BEC crossover in the cuprates \cite{Perali2002}. Our calculated spectral structures are not related to such effects but occur due to the completeness of our Eliashberg theory, i.e.\ by taking into account multiple bands, momentum and frequency dependence, and chemical potential renormalization throughout the full electron bandwidth.

\begin{figure}[ht!]
\hspace*{-0.3cm}
\includegraphics{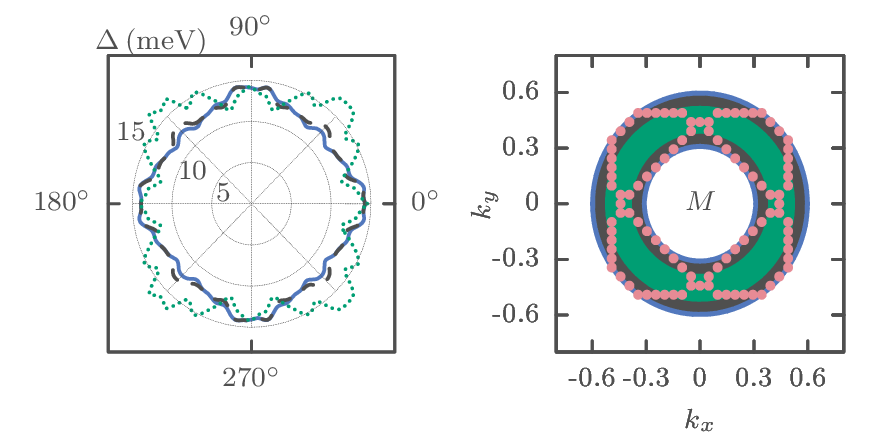}
\caption{Simulation of momentum-dependent gap measurements resulting from symmetrized EDC curves in a $20$\,meV window and for $T$=10\,K. The red dots in the right-hand graph represent Fermi surface points. The experimentally measured Fermi surface is modeled by circles of various thicknesses, over which the results are averaged. Using the same color code for both panels, the resulting gap $\Delta$ at a particular angle in the left plot is calculated by an average over all ${\bf k}$-points at this angle on the corresponding circle in the right panel. Depending on the thickness of the circle, reflecting the quality of the approximation with respect to the `real' Fermi surface, the maximum gap value is obtained at different angles and the degree of anisotropy changes. The polar axes in both panels are the same; a $0^{\circ}$ ($45^{\circ}$) angle corresponds to the line along the $X-M$ ($\Gamma-M$) direction.}
\label{polar}
\end{figure}

An experimental quantity that can be extracted from the energy location of the coherence peak in the measured EDCs is the value of the superconducting gap. The momentum dependence of the latter can then be obtained by combining EDCs from different Fermi surface momenta. This procedure not only depends on the way the Fermi surface is sampled but also on the window around the Fermi level where the spectra are integrated over energy. What is more, this procedure may be significantly complicated when the superconducting gap function is strongly momentum and energy dependent and/or when the material's electronic band structure includes shallow bands, as discussed above. For example, Lee \textit{et al.} sample the momenta used for the gap measurements from a Fermi surface that has the shape of a circular band, see Fig.\,1{(a)} in Ref.\,\cite{Lee2014}. We mimic this situation in Fig.\,\ref{polar} by testing different thicknesses of such bands in momentum space. For a given ${\bf k}$, which lies on this broadened circle, we use our symmetrized ARPES data to simulate a peak-to-peak measurement procedure for extracting the superconducting gap. Taking the average over all such momenta corresponding to a specific angle, we can produce a polar plot similar to Fig.\,2{(f)} in Ref.\,\cite{Lee2014}; it is shown in the left panel of Fig.\,\ref{polar}.

Note that the absolute value of the obtained gap $\Delta$ is not to be compared directly to the work of Lee and coworkers, since we most probably have a different electron filling. It is, however, evident that the angle of the maximal value of $\Delta$ depends on the thickness of the circular band from which we are sampling the momenta. The largest BZ area is used for the solid dark blue curve in the left panel of Fig.\,\ref{polar}, which results in a maximum gap at $0^{\circ}$ (i.e. along the $X-M$ symmetry line of the folded BZ). On the contrary, choosing the smallest thickness yields a maximum $\Delta$ at $45^{\circ}$ (i.e. along the $\Gamma-M$ direction), shown by the green dotted line. The most isotropic gap, qualitatively comparable to results reported in Ref.\,\cite{Lee2014}, is shown by the dashed black line. Interestingly, the case where the gap maxima lie along the $\Gamma-M$ direction of the BZ (green dotted line) is in good agreement with more recent ARPES measurements \cite{Zhang2016}. There it was shown that this particular gap anisotropy cannot be fitted by assuming different form factors for the symmetry of the gap and the possibility of a sign alternating gap or a competition between intra- and interorbital pairing was suggested. In contrast, here we find a similar gap anisotropy with an anisotropic s-wave gap which is driven by the concomitant momentum decoupling of the small-$\bf q$ interfacial EPI \cite{Varelogiannis1996}. These simulations lead to the conclusion that the gap measurement in this material strongly depends on both the Fermi-surface sampling and the measurement angle. Caused by the possibly large anisotropy of $\Delta$, deviations of more than 5 meV (see the green dotted line in Fig.\,\ref{polar}) are possible for a fixed Fermi surface approximation, depending only on the angle.

\begin{figure}[t!]
\includegraphics[width = 1.0\columnwidth, clip, unit=1pt]{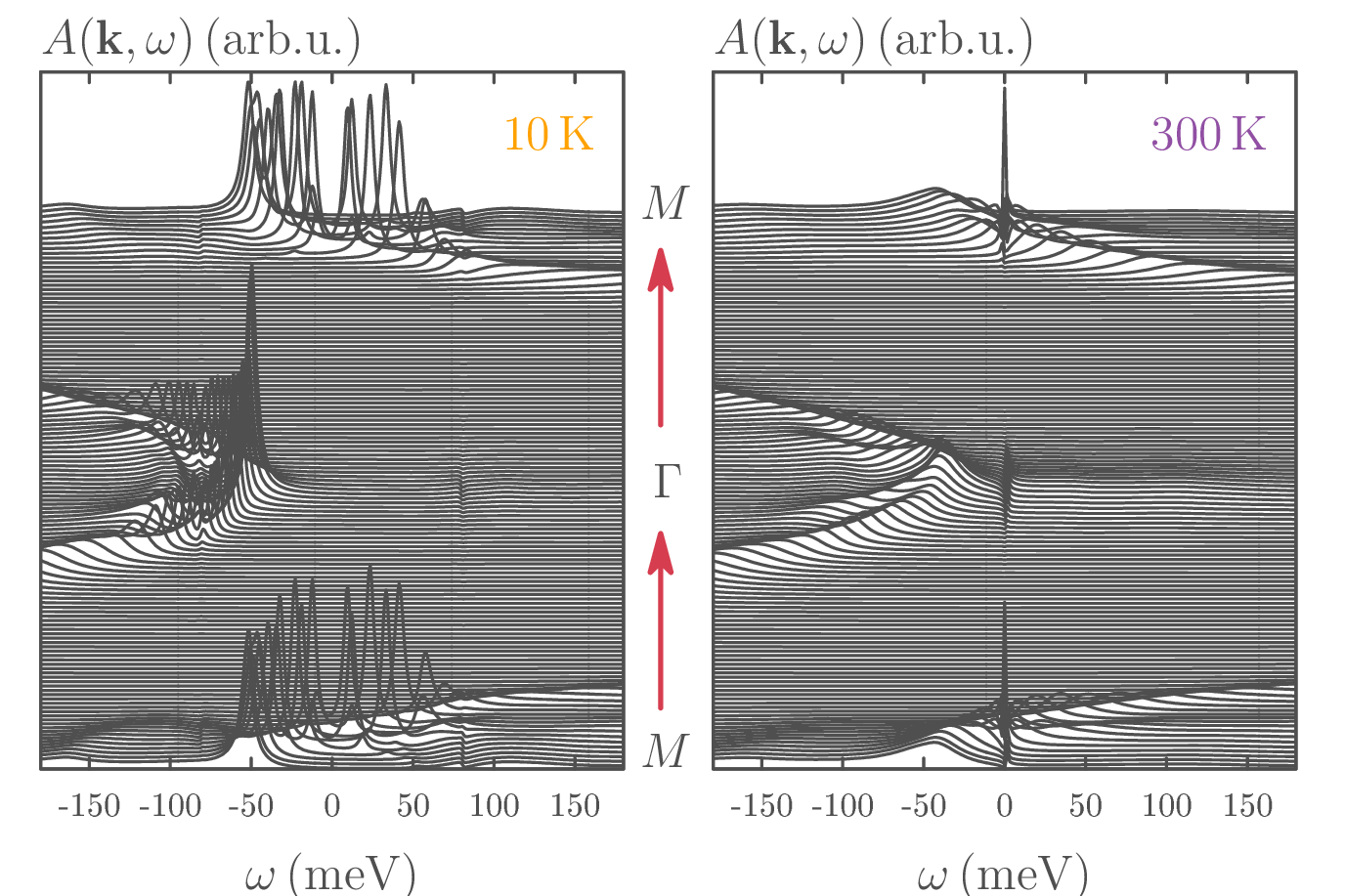}
\caption{The spectral function $A({\bf k},\omega)$ calculated as a function of energy along the high-symmetry line $M-\Gamma-M$, shown for temperatures $T=10$\,K (left) and $T=300$\,K (right). Apart from the hole bands at $\Gamma$ and the electron bands at $M$, the superconducting coherence peaks are clearly visible in the left panel. In addition, in the left panel the phonon kink appears precisely at the characteristic mode of $81$\,meV, which we used as input for the calculations. This feature is, although mediated by electron-phonon interaction, not robust with temperature, as we observe in the right panel. Instead another peak appears at small positive energies (compare also Fig.\,\ref{arpes}).} 
\label{arpfea} 
\end{figure}

Due to the efficacy of our theory we can show the self-consistently calculated ARPES-resolvable spectral function, being a function of energy, along the high-symmetry line $M-\Gamma-M$ in the first Brillouin zone for temperatures below and above $T_c$, see Fig.\,\ref{arpfea}. It is easily observed that in the superconducting state (left panel) coherence peaks appear at the electron bands at $M$ and sharp peaks appear at the hole bands at $\Gamma$; after the transition to the normal state (right) the coherence peaks vanish and the remaining quasiparticle peaks are less pronounced and more spread.  In the same Figure one can also discern the formation of the replica bands near the $M$ point. Further examination reveals that the characteristic phonon peak present for $10$\,K at an energy of $81$\,meV is being washed out due to thermal effects. As we stated already when describing Fig.\,\ref{arpes}, this thermal broadening results in a signal developing slightly above zero frequency; due to this feature appearing only near the main energy bands, our explanation of this being caused by Fermi surface spectral weight spreading is well justified.

\section{Conclusions}

We have presented the first self-consistent full bandwidth, multiband and anisotropic Eliashberg theory with inherent temperature dependence and fixed doping level for the FeSe monolayer on the STO substrate. The developed procedure is generally applicable to other materials and serves as a way to distinguish doping from temperature effects. Within our treatment we observed that increasing temperature from 10 to 300\,K leads to a 5\,meV shift of the global chemical potential. This shift is less than what was recently reported for bulk FeSe, but it may be resolvable in future experiments. Moreover, we found a non-trivial behavior of the global chemical potential below $T_c$ that should be accessible in experiments, and which indicates that superconductivity in FeSe/STO is not far from the BCS-BEC crossover regime, similarly to the situation in bulk FeSe. Regarding the latter, it has been shown recently that this regime can be accessed by either applying magnetic fields \cite{Kasahara2014} or doping \cite{Rinott2017}. Hence, it would be worth investigating whether FeSe/STO can be tuned in a similar manner through the BCS-BEC crossover regime. Further, we were able to observe an approximately decoupled energy and temperature dependence for the momentum-averaged chemical potential renormalization and $\mu$, respectively. Though not yet measured, with raising temperature we can predict both, an almost constant Fermi surface and, to good approximation, fixed positions of the main and replica electronic energy bands at the $M$ point that should be observable in ARPES measurements. Additionally, we observe the formation of second-order replica bands for which we highlight here their potential importance in accurately determining the energy scale of the interfacial phonon, when measured in conjunction with the main replica bands. For temperatures well above $T_c$ we find another peak developing at energies slightly above zero. We could furthermore show that the generally accepted symmetrization method in ARPES measurements remains trustworthy with respect to how the gap is determined, but it can introduce large biases when focusing on features other than the position of the superconducting coherence peaks. Difficulties in measuring $\Delta$ can, however, nevertheless occur if the Fermi surface sampling is inaccurate, or the superconducting gap function is very anisotropic and/or energy dependent.

The origin of the high-temperature superconductivity in FeSe/STO has been debated and attributed to conventional as well as unconventional mechanisms \cite{Aperis2011,Lee2014,Fan2015,1367-2630-17-7-073027,Rademaker2016,Linscheid2016a,Aperis2018}. Our self-consistent multiband Eliashberg-theory calculations provide results which support the picture of phonon-mediated superconductivity in FeSe/STO. This does not exclude the possible presence of spin fluctuations, but suggests that these are not primary  to the superconductivity. We do note, however, that very recent scanning tunneling spectroscopy experiments have observed a ``dip-hump" structure which has been interpreted as a possible signature of a magnetic excitation \cite{Jandke2017,Liu2018}, soliciting thus further studies of the origin of the tunneling spectrum of this remarkable system.

On a more general note, our extension of the full bandwidth, multiband and anisotropic Eliashberg theory to systematically include temperature dependence while self-consistently accounting for the chemical potential, opens up perspectives for future fully \textit{ab initio} calculations of phonon \cite{Aperis2015} and spin-fluctuation \cite{Bekaert2018} mediated superconductivity as well as of  concomitant electronic band renormalizations.

\begin{acknowledgments}
	This work has been supported by the Swedish Research Council (VR), the R{\"o}ntgen-{\AA}ngstr{\"o}m Cluster, and the Swedish National Infrastructure for Computing (SNIC).	
\end{acknowledgments}

\appendix

\section{Anisotropic multiband Eliashberg theory\label{app1}}

In this Appendix we provide details about the coupled set of equations which we solve in Matsubara space to obtain the results shown in the main text. The microscopic Hamiltonian of the system which we consider consists of a phonon, an electronic and a coupling part, where the electrons are assumed to originate in the FeSe layer and the phonons come from the substrate. By using $\hat{\rho}_i$, $i = 0, \cdots , 3$, as the usual set of Pauli matrices and $\xi^b_n({{\bf k}})$ as the band-dependent electronic energy dispersion, which we obtain from a ten-band tight-binding model based on Density Functional Theory, we can write
\begin{eqnarray}
H&=&\sum_{{\bf k},n}\left(\xi^b_n ({\bf k})-\mu\right) \Psi_{{\bf k}n}^{\dagger}\hat{\rho}_3\Psi_{{\bf k}n}^{ } + \sum_{{\bf q}}\hbar\Omega\left(b_{{\bf q}}^{\dagger}b_{{\bf q}}+\frac{1}{2}\right) \nonumber \\
&&+ \sum_{{\bf k},{\bf k}^{\prime}}\sum_{n,n^{\prime}}g_{{\bf q}}^{nn^{\prime}}u_{\bf q}\Psi_{{\bf k}^{\prime}n}^{\dagger}\hat{\rho}_3\Psi_{{\bf k}n^{\prime}}^{~}  ~~,
\label{hamiltonian}
\end{eqnarray}
with $\Psi^\dagger_{{\bf k}n}=(c_{{\bf k}\uparrow,n}^{\dagger},~c_{-{\bf k}\downarrow,n})$ the electron Nambu spinors. The creation and annihilation operators are denoted as $c_{{\bf k}n}^{\dagger},~c_{{\bf k}n}^{ }$ and $b_{{\bf q}}^{\dagger},~b_{{\bf q}}$ for fermions and bosons, respectively, with $n$ the band index and ${\bf k}$, ${\bf q}$ momentum vectors. The displacements of the phonons in Eq.\,(\ref{hamiltonian}) are defined as $u_{{\bf q}}$, the Einstein-like phonon frequency of the optical mode is given by $\Omega$. We assume the electron-phonon coupling to be band-independent, i.e.\ $g_{{\bf q}}\equiv g_{{\bf q}}^{nn^{\prime}}$, and define it to have the functional form $g_{{\bf q}}=g_0\exp(-a|{\bf q}|/0.3)$ \cite{Lee2014}; $a$ is the  FeSe lattice constant and $g_0$ the global effective EPI strength. In the definition of the Hamiltonian (\ref{hamiltonian}) the Coulomb interaction is not explicitly accounted for (see discussion in Ref.\,\cite{Aperis2018}). We treat the electron self-energy in the Migdal limit, which has been shown to remain a valid approximation even under non-adiabatic conditions when the electron-phonon interaction is dominated by forward scattering \cite{Abrikosov2005,Wang2016c}. Further following the standard Eliashberg treatment we find the electronic self-energy as 
\begin{eqnarray}\no
\hat{\Sigma}_n({\bf k}, i\omega_m) = T\sum_{{\bf k}^{\prime},m^{\prime}}\sum_{n^{\prime}}\hat{\rho}_3\hat{G}_{n^{\prime}}({\bf k}^{\prime}, i\omega_{m^{\prime}})\hat{\rho}_3 \\ \times\int_0^{\infty}\text{d}\omega\frac{\alpha^2F_{nn^{\prime}}({\bf k},{\bf k}^{\prime}; \omega)}{N_{n^{\prime}}(0)}\frac{2\omega }{(\omega_m-\omega_{m^{\prime}})^2 + \omega^2} ,
 \label{selfenergy} 
\end{eqnarray}
with temperature $T$, fermionic Matsubara frequencies $\omega_m=\pi T(2m+1)$ and the band, momentum and frequency dependent matrix Green's function is defined as
\begin{eqnarray}\no
\hat{G}_n({\bf k}, i\omega_m)=\Bigl[i\omega_mZ({\bf k},i\omega_m)\hat{\rho}_0 - \phi({\bf k},i\omega_m)\hat{\rho}_1\\
- \left[\xi^b_n({\bf k})-\mu+\chi({\bf k},i\omega_m)\right]\hat{\rho}_3 \Bigl]\Theta^{-1}_n({\bf k},i\omega_m) ,\label{greensfun}
\end{eqnarray}
with
\begin{eqnarray}
\Theta_n({\bf k},i\omega_m) &=& -\left[\omega_mZ({\bf k},i\omega_m)\right]^2 - \phi^2({\bf k},i\omega_m) \label{couplednorm} \nonumber \\
&& - \left[\xi_n^b({\bf k})-\mu+\chi({\bf k},i\omega_m)\right]^2 .
\end{eqnarray}

The self-energy (\ref{selfenergy}) contains the Density of States at the Fermi level $N_{n^{\prime}}(0)$ and the Eliashberg function,
\begin{eqnarray}
\alpha^2F_{nn^{\prime}}({\bf k},{\bf k}^{\prime}; \omega) &\equiv& \alpha^2F_{n'}({\bf k},{\bf k}^{\prime}; \omega) \nonumber \\
&=& N_{n^{\prime}}(0)|g_{\bf q}|^2\delta(\omega-\Omega) .
\label{eliashfun}
\end{eqnarray}

Within Eliashberg theory, we obtain the set of three coupled and self-consistent equations, describing the mass renormalization $Z$, the gap function $\phi$ and the chemical potential renormalization $\chi$; these are given below in Eqs.\,(\ref{coupledfun1})-(\ref{coupledfun3}). The Matsubara frequency and momentum dependent electron-phonon interaction is defined by an integral over real frequencies of the Eliashberg function as
\begin{eqnarray}
V_{e-ph}({\bf q},\omega_m-\omega_{m^{\prime}}) = ~~~~~~~~~~~~~~~~~~~~~~~~~~~~~~~~~&&  \\ \int_0^{\infty}\text{d}\omega \frac{\alpha^2F_{n'}({\bf k},{\bf k}^{\prime}; \omega)}{N_{n^{\prime}}(0)} \frac{2\omega }{(\omega_m-\omega_{m^{\prime}})^2 + \omega^2} && ~~, \nonumber
\end{eqnarray}
which directly comes from Eq.\,(\ref{selfenergy}). The superconducting gap function can be found by the familiar expression $\Delta({\bf k},i\omega_m)=\phi({\bf k},i\omega_m)/Z({\bf k},i\omega_m)$. The quantities in Eqs.\,(\ref{coupledfun1})-(\ref{coupledfun3}) are to be solved iteratively in coupled momentum and Matsubara space.
  As described in the main text we extend the treatment by an additional equation for the chemical potential, with which we are able to keep the electron filling at a desired level. This procedure, in particular, allows us to isolate the temperature dependence of various quantities. Although it is not an easy task to couple Eq.\,(\ref{mu}) to $Z$, $\chi$, and $\phi$, there is neither a significant increase in the computational complexity of the algorithm, nor is there a need for a much larger number of iterations or Matsubara frequencies to get the desired precision. To derive Eq.\,(\ref{mu}) we split the Matsubara sum of Eq.\,(\ref{n1}) into a normal-state and a superconducting part, as described in the main text. This normal-state expression can be evaluated analytically and gives a finite correction to the `usual' sum bounded by a hard cutoff.

\begin{eqnarray}\no
Z({\bf k}, i\omega_m) &=& 1 - \frac{T}{\omega_m}\sum_{{\bf k}^{\prime},m^{\prime}}\sum_n V_{e-ph}({\bf q},\omega_m-\omega_{m^{\prime}})\\\label{coupledfun1}
&\times&\frac{\omega_{m^{\prime}}Z({\bf k}^{\prime},i\omega_{m^{\prime}})}{\Theta_n({\bf k}^{\prime},i\omega_{m^{\prime}})}  \\ \nonumber
\chi({\bf k}, i\omega_m) &=&  T\sum_{{\bf k}^{\prime},m^{\prime}}\sum_n V_{e-ph}({\bf q},\omega_m-\omega_{m^{\prime}})\\\label{coupledfun2}
&\times&\frac{\xi_n^b({\bf k})-\mu+\chi({\bf k}^{\prime},i\omega_{m^{\prime}})}{\Theta_n({\bf k}^{\prime},i\omega_{m^{\prime}})}  \\\no
\phi({\bf k}, i\omega_m) &=& - T\sum_{{\bf k}^{\prime},m^{\prime}}\sum_n V_{e-ph}({\bf q},\omega_m-\omega_{m^{\prime}})\\\label{coupledfun3}
&\times&\frac{\phi({\bf k}^{\prime},i\omega_{m^{\prime}})}{\Theta_n({\bf k}^{\prime},i\omega_{m^{\prime}})} ~~, 
\end{eqnarray}

The Eliashberg equations, Eq.\,(\ref{coupledfun1})-(\ref{coupledfun3}), together with the expression for the chemical potential, Eq.\,(\ref{mu}), constitute a set of four coupled equations which has been implemented in the UppSC code \cite{UppSC}. As input for our self-consistent calculations we use a parametrized ten-band set of energy dispersion $\xi_n^b ({\bf k})$ that have been obtained from \textit{ab initio} Density Functional Theory calculations ~\cite{Eschrig2009} for bulk FeSe and have been adjusted to the monolayer case, as discussed in Ref.\,\cite{Hao2014}. When deposed on the substrate, the lattice constant of the monolayer is distorted, which has to be taken into account. An important feature of the energy dispersions used for our calculations is that only two bands are crossing the Fermi level near the $M$ point. The hole bands located at $\Gamma$ are below the Fermi energy.  The bands at $M$ have been shown to yield the largest contributions to several, but not all superconducting properties \cite{Aperis2018}. These are the bands that have been used for creating the Fermi surfaces shown in Fig.\,\ref{3dFSplot}.

Within our numerical algorithm we use the strict convergence criterion of $10^{-8}$ for the relative error for each function and for all momenta and energies. We are even able to keep the electron filling constant with respect to the initial value, almost up to numerical accuracy. As described in the main text, for the calculations we used the constants $g_0=728$ meV and $\hbar\Omega=81$ meV \cite{Aperis2018}. To be confident about well converged results in Matsubara space, we chose about 3000 frequencies, which corresponds to approximately twice the full electronic band width.

\begin{figure}[bh]
\includegraphics{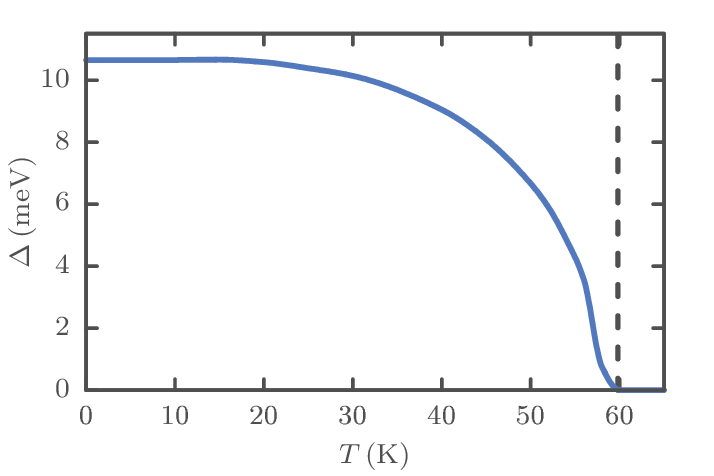}
\caption{Calculated maximum superconducting gap edge among Fermi surface momenta. The gap follows the usual trend leading to $T_c\sim60$\,K, indicated by the dashed gray line.}
\label{gpdmax}
\end{figure}
\section{From Matsubara to real frequencies\label{app2}}

Since we want to compare our theoretical results to experimental data from ARPES measurements, we need to analytically continue the results obtained in Matsubara space. Making use of the formally exact procedure brought forward in Ref.\,\cite{Marsiglio1988}, we derive another set of self-consistent coupled equations, now on the real-frequency axis, these are given below in Eq.\,(\ref{an1}-\ref{an3}). From Eq.\,(\ref{eliashfun}) we recall that the Eliashberg function contains a delta peak at the phonon frequency, which is why we introduce a Lorentzian shaped function, which is properly normalized and introduces a natural (physical) broadening \cite{Marsiglio1988,Degiorgi1994}. In addition we find that the zero-frequency components  introduce numerical instabilities if not treated with special care. For this part of our algorithm we have used the convergence criterion of a relative error $10^{-6}$ and cross-checked the results with our previous work \cite{Aperis2018} and with converged Pad\'e approximants.
\begin{widetext}
\begin{eqnarray}\no
Z({{\bf k},\omega})&=&1 -\frac{1}{\omega}{\rm T}\sum_{{\bf k'},m'}\sum_{n}V_{e-ph}({\bf q},\omega-\omega_{m'}) \frac{Z({{\bf k'},i\omega_{m'}})i\omega_{m'}}{\Theta_n({{\bf k'},i\omega_{m'}})}\\\label{an1}
&-&\frac{1}{2\omega}\int_{-\infty}^\infty dz' \sum_{{\bf k'}}\sum_{n} \frac{\alpha^2F_n({\bf k},{\bf k}^{\prime}; z^{\prime})}{N_{n}(0)}\left(\tanh{\frac{\omega-z'}{2{\rm T}}}+\coth{\frac{z'}{2{\rm T}}}\right) \frac{Z({{\bf k'},\omega-z')}(\omega-z')}{\Theta_n({{\bf k'},\omega-z'})} , \\\no
\chi({\bf k},\omega)&=&{\rm T}\sum_{{\bf k'},m'}\sum_{n}V_{e-ph}({\bf q},\omega-\omega_{m'}) \frac{\xi^b_n({\bf k'})-\mu + \chi({\bf k'},i\omega_{m'})}{\Theta_n({\bf k'},i\omega_{m'})}\\\label{an2}
&+&\frac{1}{2}\int_{-\infty}^\infty dz' \sum_{{\bf k'}}\sum_n \frac{\alpha^2F_n({\bf k},{\bf k}^{\prime}; z^{\prime})}{N_{n}(0)}\frac{\xi^b_n({\bf k'})-\mu + \chi({\bf k'},\omega-z')}{\Theta_n({\bf k'},\omega-z')}\left(\tanh{\frac{\omega-z'}{2{\rm T}}}+\coth{\frac{z'}{2{\rm T}}}\right) , \\\no
\phi({\bf k},\omega)&=&-{\rm T}\sum_{{\bf k'},m'}\sum_n V_{e-ph}({\bf q},\omega-\omega_{m'}) \frac{\phi({\bf k'},i\omega_{m'})}{\Theta_n({\bf k'},i\omega_{m'})}\\\label{an3}
&-&\frac{1}{2}\int_{-\infty}^\infty dz' \sum_{{\bf k'}} \sum_{n}\frac{\alpha^2F_n({\bf k},{\bf k}^{\prime}; z^{\prime})}{N_{n}(0)} \frac{\phi({\bf k'},\omega-z')}{\Theta_n({\bf k'},\omega-z')}\left(\tanh{\frac{\omega-z'}{2{\rm T}}}+\coth{\frac{z'}{2{\rm T}}}\right).
\end{eqnarray}
\end{widetext}

The superconducting gap edge at the Fermi level is found from
\begin{eqnarray}
\Delta^2({{\bf k}_{\rm F},\omega})=\omega^2 - \left[\frac{\xi^b_n({{\bf k}_{\rm F}})-\mu + \chi({{\bf k}_{\rm F},\omega})}{Z({{\bf k}_{\rm F},\omega})}\right]^2 , ~~~~
\end{eqnarray}
with the usual convention $\Delta=\phi/Z$. The calculated maximum of the superconducting gap edge is shown in Fig.\,\ref{gpdmax} as a function of temperature. It is easily seen that with a temperature resolution of 1\,K we obtain a superconductivity transition temperature of about 60\,K, which agrees very well with experiment \cite{Lee2014}. Making use of the fact that $\Delta=0$\,eV above $T_c$, and considering the zero frequency component, we find the Fermi surface from the condition $\xi^b_n({{\bf k}_{\rm F}})-\mu + \text{Re}(\chi_0({{\bf k}_{\rm F}},0))=0$, with the index 0 of $\chi$ denoting the temperature above $T_c$ \cite{Aperis2018}. Note, that as we describe in the main text, the Fermi surface is in general temperature dependent, though the variations are found to be very small. 

\bibliographystyle{apsrev4-1}
\end{document}